\documentclass[prb,superscriptaddress,twocolumn]{revtex4}
\usepackage{epsfig,amsmath,amssymb,latexsym}
\usepackage{graphicx}
\usepackage{dcolumn}
\usepackage{bm}
\usepackage{subfigure}

\begin{document}


\title{Screening in graphene antidot lattices}

\author{M. H. Schultz}
\affiliation{Department of Micro and Nanotechnology, DTU Nanotech, Bldg. 345B,
Technical University of Denmark, DK-2800 Kongens Lyngby, Denmark}
\author{A.~P. Jauho}
\affiliation{Department of Micro and Nanotechnology, DTU Nanotech, Bldg. 345B,
Technical University of Denmark, DK-2800 Kongens Lyngby, Denmark}
\affiliation{Department of Applied Physics, Helsinki University of
Technology, P.O.Box 1100, FI-02015 TKK, Finland}\author{T.~G.
Pedersen} \affiliation{Department of Physics and Nanotechnology,
Aalborg University, DK-9220 Aalborg East, Denmark}

\date{\today}

\begin{abstract}
We compute the dynamical polarization function for a graphene
antidot lattice in the random-phase approximation.  The computed
polarization functions display a much more complicated structure
than what is found for pristine graphene (even when evaluated beyond
the Dirac-cone approximation); this reflects the miniband structure
and the associated van Hove singularities of the antidot lattice.
The polarization functions depend on the azimuthal angle of the {\bf
q}-vector. We develop approximations to ease the numerical work, and
critically evaluate the performance of the various schemes.  We also
compute the plasmon dispersion law, and find an approximate
square-root dependence with a suppressed plasmon frequency as
compared to doped graphene. The plasmon dispersion is nearly
isotropic, and the developed approximation schemes agree well with
the full calculation.
\end{abstract}

\pacs{Valid PACS appear here}
\maketitle

\section{\label{sec:Intro}Introduction}

Graphene Antidot Lattices (GALs) have been promoted as a
flexible platform for creating a tunable band gap thereby possibly
allowing the realization of a number of technological applications
\cite{Pedersen_PRL_2008}. In its basic form the GAL consists of a graphene sheet
with a periodic nanometer scale perforation.  Alternatively, GALs can
be viewed as a special realization of a superlattice imposed on
pristine graphene. Such a superlattice can be fabricated by a variety
of technologies, for example by a periodic chemical modification, by selective
adsorption of atoms or molecules on graphene, by periodic
arrangement of electrostatic gates, or by an intrinsic or extrinsic
regular corrugation. There is already a substantial literature on
this general topic (see, e.g., Refs.
[\onlinecite{Park_NPhys_2008,Bai_PRB_2007,Barbier_PRB_2008,Isacsson_PRB_2008,Balog_NMat_2010,Brey_PRL_2009}]);
here we focus on the antidot lattice case. Theoretically the
properties of triangular GALs have been examined already quite
substantially (e.g., optical properties \cite{Pedersen_PRB_2008},
excitons \cite{Pedersen_PRB_2009}, electronic properties
\cite{Furst_NJP_2009,Vanevic_PRB_2009}, electron-phonon coupling
\cite{Vukomirovic_PRB_2010,Stojanovic_PRB_2010}, detection of edge
states \cite{Wimmer_PRB_2010}, or details of band gap scaling
\cite{Liu_PRB_2009}). Most importantly, the experimental techniques
for fabricating GALs have evolved rapidly
\cite{Eroms_NJP_2009,Bai_NNano_2010,Kim_NanoLett_2010,Beg_NanoLett_2011}, presently
reaching lattice constants down to a few tens of nanometers, where
many of the interesting quantum mechanical effects predicted by
theory should become detectable.  Of particular interest are transport
measurements, which we expect soon to be available.

Most of the early works on GALs have focused on single-particle
effects, such as the miniband structure, sometimes supported by {\it
ab initio} calculations using density-functional theory (DFT), or by
some other less accurate but computationally more effective method,
such as the density-functional based tight-binding (DFTB)
\cite{Furst_PRB_2009}. However, it is necessary also to consider
electron-electron interactions since these affect many physical
properties, such as the charge carriers' interaction with light or
with lattice vibrations, dielectric screening and plasmons, or the
transport properties. The dynamical polarization function is a
central object in these considerations, and it has been a topic of
wide interest. In the literature expressions relevant for the
polarization function of graphene, evaluated in the random phase
approximation, can be found already from the "pre-graphene" era
(see, e.g., Refs.~[\onlinecite{Shung_PRB_1986,Gonzales_PRB_1999}]).
More recently, thorough studies of the polarization function in the
Dirac-cone approximation have been reported in
Refs.~[\onlinecite{Wunsch_NJP_2006,Hwang_PRB_2007}], but only very
recently Stauber et al. \cite{Stauber_PRB_2010,Stauber_PRB_RC_2010}
gave similar results for full graphene dispersion at finite chemical
potential. It should be noted that the accuracy of the RPA for
pristine graphene is a highly nontrivial issue, see, e.g.,
Ref.~[\onlinecite{Abedinpour_preprint}], and references cited
therein.  Many of the complications encountered in these studies
originate from the high-momentum cutoff needed in the Coulomb
interaction, when extending the discussion beyond RPA, and are
specific to the Dirac cone dispersion law. Here, we consider the
full GAL dispersion with a gap, and believe that the RPA is a
justifiable starting point.  Our calculations are conceptually
straightforward: indeed the formal expressions for the Lindhard
function for a general tight-binding model (which will be our
underlying antidot lattice Hamiltonian) are standard text-book
material (see, e.g., Sect. 8.5 in Ref.~[\onlinecite{Bruus_2004}]).
Some complications inevitably arise because the antidot
lattice unit cells contain tens or hundreds of atoms, in contrast to
just two in pristine graphene. Analytic expressions can hardly be
expected for such a complicated system, and much of our effort goes
into examining whether simpler yet reasonably accurate models can reproduce
the results of the full numerical calculations.  The GALs can be
realized with many different lattice symmetries, each of which may
have some interesting physics of its own (see, e.g.
Refs.~[\onlinecite{Vanevic_PRB_2009,Petersen_ACSNano_2011,Ouyang_ACSNano_2011}]):
whether a gap arises in the energy spectrum depends in a sensitive way on the
symmetry of the antidot lattice, and on the orientation of the antidot lattice
with respect to the underlying graphene lattice. Here
we examine one particular GAL (with a triangular lattice symmetry) in detail as a demonstration of the
theoretical methodology, and defer the discussion of other lattice
symmetries until experiments performed on well-characterized samples
become available.

This paper is organized as follows.  In Sect.~\ref{sec:model} we
briefly introduce the basic models used in this work: the
tight-binding representation of the antidot lattice states, and a
simplified model, the "gapped graphene". In
Sect.~\ref{sec:results_pol} we give both analytical and numerical
results for the polarization function, address certain issues in the
numerical computation, and present a comparison between the various
approximation schemes. Sect.~\ref{sec:plasmons} addresses the
plasmonic properties of GALs, and we end with a brief conclusion.

\section{\label{sec:model}The models}
\subsection{\label{subsec:energy}Antidot lattice energy spectrum}
\begin{figure}[htp]
\includegraphics[width=8.5cm] {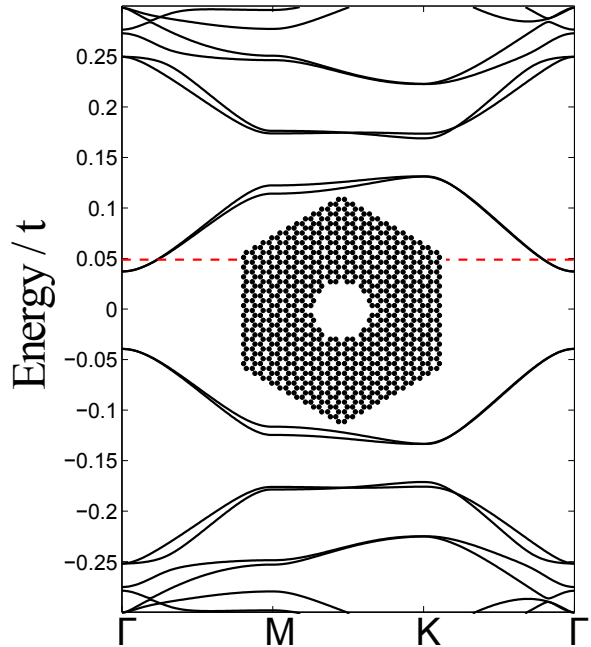}
\caption{(Color online) Band structure of a $\{12,3\}$ GAL with a circular antidot with an armchair edge.
The inset shows the atomistic configuration of the unit
cell. The dashed line indicates the value of the chemical potential,
$\mu=0.05 t$ ($t$ is the hopping integral), used in most of the
subsequent calculations. \label{fig:energies}}
\end{figure}

We use a tight-binding model to compute the band structure of the
GALs. The wave function is written as
\begin{equation}
\psi_{n{\bf k}}({\bf r})=\frac{1}{\sqrt{N}}\sum_{\bf R}e^{i{\bf
k}\cdot{\bf R}}\sum_{j=1}^M c_{n{\bf k}}^j \phi({\bf r}-{\bf R}-{\bf
d}_j),\label{tbwavefunction}
\end{equation}
where $\phi({\bf r})$ is the normalized atomic wave function of the
$2p_z$ orbital of carbon, and the $j$-sum runs over the atoms in
the unit cell of the antidot lattice. {\bf R} runs over the unit
cells in the antidot lattice. The coefficients $c_{n{\bf k}}^j$ and
the eigenenergies $\epsilon_{n{\bf k}}$ are then obtained from the
$M\times M$ eigenvalue problem, for each value of the wave vector
$\bf k$ in the 1st Brillouin zone of the antidot lattice. Each GAL is
characterized by a pair of indices $\{L,R\}$, where the $L$ index
gives the side length of the hexagonal unit cell, and the $R$
index is the (approximate) radius of the hole, in units of the
graphene lattice constant $\sqrt{3}a$, where $a=1.42$ {\AA}  is the C-C
distance\cite{Pedersen_PRL_2008}. Figure \ref{fig:energies} shows an example of a typical GAL band structure.
\subsection{\label{subsec:gap}Gapped graphene}
The gapped graphene model is a phenomenological model used to
describe band structures with a gap. Consider first the pristine
graphene dispersion $\epsilon_{n{\bf k}}^0=nt|\phi({\bf k})|$, where
$t=3.03$ eV is the hopping integral, $n=\pm 1$ and
\begin{equation}
\phi({\bf k})=e^{i{\bf k\cdot\boldsymbol\delta_1}}+e^{i{\bf
k\cdot\boldsymbol\delta_2}}+e^{i{\bf k\cdot\boldsymbol\delta_3}},\label{phi}
\end{equation}
where $\boldsymbol\delta_i$ are the three neighbors of the reference
atom at the origin ($\boldsymbol\delta_1=a/2(-1,\sqrt{3})$,
$\boldsymbol\delta_2=a/2(-1,-\sqrt{3})$, $\boldsymbol\delta_3=a(1,0)$).
Next, introduce a shift $\pm\Delta$ on the onsite-energies for the
carbon atoms on the $A$- and $B$- sublattices, respectively. This
gives rise to a new spectrum,
\begin{equation}
\epsilon_{n{\bf k}}=n\sqrt{\Delta^2+t^2|\phi_{\bf
k}|^2},\label{gappedenergy}
\end{equation}
where $n=\pm 1$, and corresponding eigenvectors
\begin{equation}{\bf v}_{n{\bf k}}=\frac{1}{\sqrt{2}} \left(
\begin{array}{c}
\frac{-nt\phi_{\bf k}}{\sqrt{|\epsilon_{n{\bf k}}|(|\epsilon_{n{\bf k}}|-n\Delta)}} \\
\sqrt{\frac{|\epsilon_{n{\bf k}}|-n\Delta}{|\epsilon_{n{\bf k}}|}}\\
\end{array}
\right)\label{gappedstate}.
\end{equation}
Since $\phi({\bf k})$ vanishes at the Dirac points $K,K'$, the
energy gap is $E_g=\epsilon_{+,{\bf K}}-\epsilon_{-,{\bf
K}}=2|\Delta|$, and $\Delta$ can be chosen so that the spectrum fits
the gap of a given antidot lattice spectrum.  The gapped graphene can
also be viewed as a minimal perturbation of the pristine graphene, e.g., in
the case of very small antidots compared to the full unit cell. If, moreover,
the chemical potential is close to the band edge so that the low-energy part of
the band structure dominates, we expect the gapped graphene model to be accurate.
Below we give explicit examples of what "low-energy" means in practice.

\section{\label{sec:results_pol}Polarization function}
\subsection{\label{subsec:analytics}Analytical results}
The polarization function is evaluated from the density-density
response function for noninteracting electrons:
\begin{eqnarray}\chi^r_0({\bf
q},\omega)&=&\frac{2}{(2\pi)^2}\sum_{nn'{\bf k}'}\int_{1.{\rm
BZ}}d{\bf k}\left|\langle n{\bf k}|e^{-i{\bf q}\cdot{\bf r}}|n'{\bf
k}'\rangle\right|^2\nonumber\\
&\quad&\times\frac{n_F(\epsilon_{n{\bf k}})-n_F(\epsilon_{n'{\bf
k}'})}{\epsilon_{n{\bf k}}-\epsilon_{n'{\bf
k}'}+\omega+i\eta}.\label{chi0}
\end{eqnarray}
Here, $n_F$ is the Fermi-Dirac distribution, which we evaluate at a fixed
temperature of $k_B T=0.01t$ throughout.
In writing Eq. (\ref{chi0}) we have assumed that local field effects can
be neglected, i.e., the relevant wave vectors $q$ are small compared
to the reciprocal lattice vectors $G\propto 1/L$. This allows us to
work with a scalar dielectric function, rather than the more general
object defined in the ${\bf G},{\bf G}'$-space.  The price is that
certain collective intervalley modes will not be included (see,
e.g., Ref. [\onlinecite{Tudor_PRB_2010}]).

In addition to the energy spectrum one also needs to evaluate the
matrix element. In the following we list the required matrix element
for the various cases of interest, including some well-known results
for completeness.
\subsubsection{\label{pristine}Pristine graphene} The
matrix element is \cite{Stauber_PRB_2010}
\begin{eqnarray}&&\left|\langle n{\bf k}|e^{-i{\bf q}\cdot{\bf r}}|n'{\bf
k}'\rangle\right|^2=\delta_{{\bf k}',{\bf k}+{\bf
q}}\nonumber\\
&\quad&\times\frac{1}{2}\left(1+nn'{\rm Re}\left[e^{i{\bf
q}\cdot{\boldsymbol\delta}_3}\frac{\phi_{\bf k}}{|\phi_{\bf
k}|}\frac{\phi_{{\bf k}+{\bf q}}^*}{|\phi_{{\bf k}+{\bf
q}}|}\right]\right),\label{fgraphene}
\end{eqnarray}
where the various quantities are defined in Eq.(\ref{phi}).
\subsubsection{\label{Dirac}Dirac cone}
With the spectrum linearized in the vicinity of the Dirac points,
\begin{equation}
\epsilon^0_{n{\bf k}}\approx n\frac{3at}{2}k,
\end{equation}
Eq.(\ref{fgraphene}) simplifies, and one
finds\cite{Wunsch_NJP_2006,Hwang_PRB_2007}
\begin{equation}\left|\langle n{\bf k}|e^{-i{\bf q}\cdot{\bf r}}|n'{\bf
k}'\rangle\right|^2=\delta_{{\bf k}',{\bf k}+{\bf
q}}\frac{1}{2}[1+nn'\cos(\theta_{{\bf k},{\bf k}+{\bf q}})],
\end{equation}
where $\theta_{{\bf k},{\bf k}+{\bf q}}$ is the angle between {\bf
k} and ${\bf k}+{\bf q}$.
\subsubsection{\label{Gapped}Gapped graphene}
Using the eigenenergies (\ref{gappedenergy}) and eigenvectors
(\ref{gappedstate}) one finds
\begin{eqnarray}&&\left|\langle n{\bf k}|e^{-i{\bf q}\cdot{\bf r}}|n'{\bf
k}'\rangle\right|^2=\delta_{{\bf k}',{\bf k}+{\bf
q}}\nonumber\\
&\quad&\times\frac{1}{2}\left(\frac{t^4|\phi_{\bf k}|^2|\phi_{{\bf
k}'}|^2+(|\epsilon_{n{\bf
k}}|-n\Delta)^2(|\epsilon_{n{\bf
k}'}|-n'\Delta)^2} {2|\epsilon_{n{\bf
k}}|(|\epsilon_{n{\bf
k}}|-n\Delta)|\epsilon_{n{\bf
k}'}|(|\epsilon_{n{\bf k}'}|-n'\Delta)}\right.\nonumber\\
&\quad&+\left.nn'{\rm Re}\left[e^{i{\bf q}\cdot{\boldsymbol
\delta}_3}\frac{t^2\phi_{\bf k}\phi_{{\bf
k}'}^*}{|\epsilon_{n{\bf k}}||\epsilon_{n{\bf
k}'}|}\right]\right).
\end{eqnarray}
\begin{figure}\includegraphics[width=8.5cm]{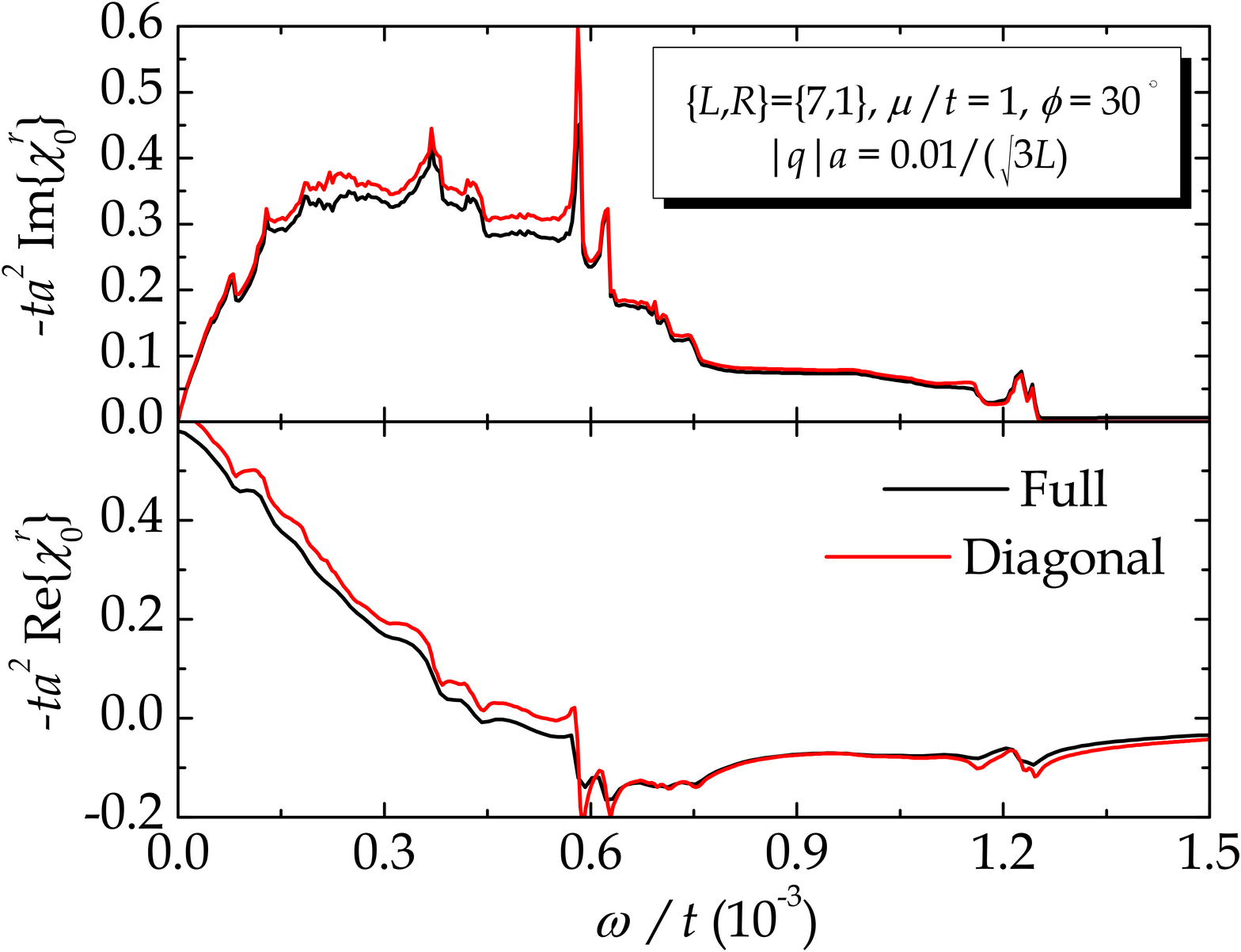}
\caption{(Color online) Im$\chi_0^r({\bf q},\omega)$ (top panel), and
Re$\chi_0^r({\bf q},\omega)$ (bottom panel)  as a function
of $\omega/t$ for a $\{7,1\}$ GAL. $\phi$ defines the azimuthal
angle of the {\bf q}-vector. The curve labeled "full" is evaluated
using the full expression Eq. (\ref{tbmatrixelement}), while the curve
labeled "diagonal" employs the approximation Eq. (\ref{wildappro}).}
\label{testIm}
\end{figure}

\subsubsection{\label{Antidot}Antidot lattice} Using the
tight-binding wave functions (\ref{tbwavefunction}) the matrix
element becomes
\begin{equation}
\langle n{\bf k}|e^{-i{\bf q}\cdot{\bf r}}|n'{\bf
k}'\rangle=\delta_{{\bf k}',{\bf k}+{\bf q}}\sum_{j=1}^M
\left(c_{n{\bf k}}^j\right)^*c_{n'{\bf k}+{\bf q}}^j e^{-i{\bf q}\cdot
{\bf d}_j}.\label{tbmatrixelement}
\end{equation}
It is readily verified that the general expression
(\ref{tbmatrixelement}) reproduces the standard results (i.e., set
$M=2$, ${\bf d}_1={\bf 0}$, and ${\bf d}_2={\boldsymbol\delta}_3$ for
pristine graphene).  The challenge in the numerical applications is
that for each {\bf q}-point a large number of terms needs to be
considered, and that  a full Brillouin zone summation is required
for every term. Below we discuss approximate methods of how to
bypass this difficulty.
\subsection{\label{subsec:numerics}Numerical results}
\subsubsection{\label{qL=0}The limit $qL\ll 1$}
As remarked above, in this work we do not consider local field
effects, i.e., we restrict our discussion to wave-vectors that are
small compared to the reciprocal lattice vectors, $qL\ll 1$ (a
similar restriction was used in
Ref.[\onlinecite{Stauber_PRB_2010}]). We may exploit this
restriction further by noting that the vectors ${\bf d}_j$ entering
the matrix element (\ref{tbmatrixelement}) satisfy $d_j < L$, and
therefore the phase-factor can be approximated as $|\exp(-i{\bf q}\cdot
 {\bf d}_j)|\simeq 1$. Further, since the wave vectors {\bf
q} in general are small, for most of the ${\bf k}$-vectors entering
the Brillouin zone summations the eigenvectors corresponding to
${\bf k}$ and ${\bf k}+{\bf q}$ do not differ much, and therefore we
can appeal to the orthonormality (in the band indices $n,n'$) of the
eigenvectors, and thereby obtain the final approximation for the
matrix elements:
\begin{equation}
\langle n{\bf k}|e^{-i{\bf q}\cdot{\bf r}}|n'{\bf
k}'\rangle\simeq\delta_{{\bf k}',{\bf k}+{\bf
q}}\delta_{n,n'}.\label{wildappro}
\end{equation}
We emphasize that since the limits of validity of the above
expression are not rigorously given, one must carefully check its
reliability in each case.  We have performed a large number of such
numerical tests, and representative results for a typical {\it worst
case} scenario are given in Fig. \ref{testIm}. The
results for the imaginary part, Fig.~\ref{testIm} (top panel) agree
qualitatively well: all major features are reproduced. For
the large $\omega$-limit, $\omega/t > 1.25$ in Fig.~\ref{testIm},
a small systematic difference is observed (barely observable within
the resolution of the figure); we have not been able to pinpoint the reason
for this tiny difference.  The results for the real part, Fig.~\ref{testIm} (bottom
panel),
were obtained by performing a Kramers-Kronig transformation of the
corresponding imaginary part.  Here we see perhaps an even better
agreement between the full numerics and the approximate result.
In general, the matrix element appears to be a smooth function of
{\bf q}, which does not introduce additional sharp features in
the polarization function.
A particularly interesting observation is that the high-energy tails of
the ${\rm Re}\chi_0^r$, which are important in the
evaluation of the plasmon spectrum, are in near quantitative
agreement, and we shall exploit this fact in Sect.~\ref{sec:plasmons} where we
analyze the plasmon dispersion laws.
\begin{figure}[t]
\includegraphics[width=8.5cm] {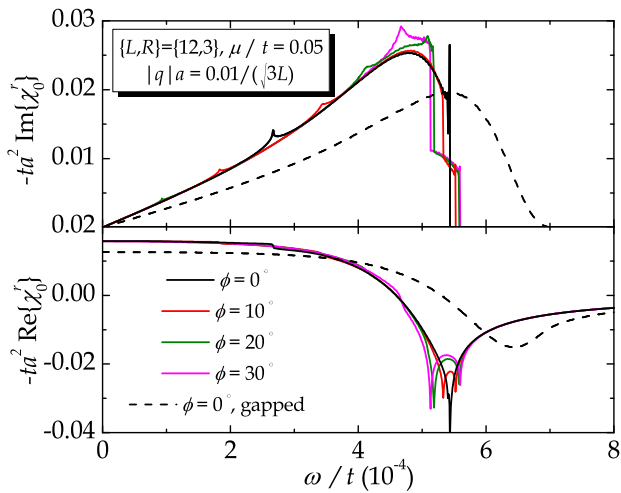}
\caption{(Color online) Im$\chi_0^r({\bf q},\omega)$ (top panel) and  Re$\chi_0^r({\bf q},\omega)$
(bottom panel) for the
$\{12,3\}$ GAL as a function of $\omega/t$, at chemical potential
$\mu/t=0.05$. The solid lines give the polarization function
(using the approximation (\ref{wildappro})) for various azimuthal
angles $\phi$ while the dashed line gives the gapped graphene
result.}\label{ImPi12_3_lowmu}
\end{figure}
\subsubsection{\label{BZ-sum}Brillouin-zone summation}
The integrals over the Brillouin zone were carried out with an
improved triangle method\cite{Pulci_PRB_1998}, see the Appendix of
Ref.~[\onlinecite{Pedersen_PRB_2008}] for details of our specific
implementation.  Very briefly, the method consists of generating a
grid of {\bf k} points, and writing the BZ-integrals as sums over
triangles spanned by the grid points, and performing an analytic
integration within each triangle using a linear approximation for
the integrand. To evaluate the integrand, the $M\times M$-eigenvalue
and -vector problem must be solved for each {\bf k} and {\bf k}+{\bf
q} pair. By checking first that the
Fermi-function difference in (\ref{chi0}) is nonzero, as well as
that the delta-function has a zero, a certain fraction of the
eigenvalue problems could be eliminated. Fortunately, the method is
rapidly converging as the {\bf k}-mesh is made denser, as could be
verified by bench-marking it against cases where analytic results
are available, e.g., the Dirac cone dispersion.  Typically we used
$\simeq 5000$ {\bf k}-points in the mesh. In our calculations we
always evaluate the imaginary part of Eq.(\ref{chi0}) first, and
then compute the real part by a Kramers-Kronig transformation.

\begin{figure}[t]
\includegraphics[width=8.5cm] {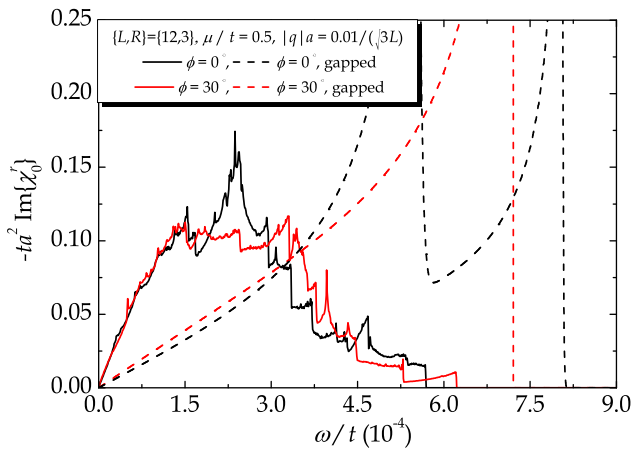}
\caption{(Color online) Imaginary part of the polarization function for the
$\{12,3\}$ GAL as a function of $\omega/t$, at chemical potential
$\mu/t=0.5$. The solid lines give the full polarization function
(using the approximation (\ref{wildappro})) for various azimuthal
angles $\phi$ while the dashed line gives the corresponding gapped
graphene results.}\label{ImPi12_3_highmu}\end{figure}
\subsubsection{\label{12-3GAL}An example: The $\{12,3\}$-GAL}
Figure \ref{ImPi12_3_lowmu}  gives the
results for the imaginary and real part of the polarization function
for the $\{12,3\}$ GAL of Fig. 1, at a relatively low chemical potential,
$\mu/t=0.05$.  We have chosen to study this particular GAL in detail,
and it is relevant to inquire how generic our results are.  Of
special interest is the effect of the shape and the edge of the
antidot, for example whether the edge is of the armchair or of
the zigzag type.  It is well-known that non-dispersive zero-energy states
may occur for certain structures (see, e.g., the examples shown in
Refs.~[\onlinecite{Vanevic_PRB_2009,Furst_PRB_2009})],
which might lead to characteristic features in $\chi_0^r({\bf q},\omega)$.  Not much
is known of the robustness of these edge states against disorder,
or other edge reconstructions; see however a recent related study by Li et al\cite{Li_Nat_Phys_2011}. A full study
of these effects is beyond the scope of the present paper, where we want to illustrate
the basic features of GALs with a relatively simple band structure, as
the one illustrated in Fig. 1.  Another attractive feature of the chosen
example is that the triangular GAL obeys a simple scaling law for
the band gap\cite{Pedersen_PRL_2008}, which implies that our results
are relevant also for larger structures, for which full numerical
calculations would be extremely challenging.

It is interesting to compare these results with those
pertaining to pristine
graphene\cite{Stauber_PRB_2010,Stauber_PRB_RC_2010}.  A certain
overall qualitative similarity persists. The imaginary part
increases as a function of the frequency, and after reaching its
maximum value it drops rapidly to zero.  Superposed are certain
sharp features: these are related to the van Hove singularities
occurring at the band edges.  Since there are many more bands in the
GALs in a given energy range than there are for pristine graphene,
it is not surprising that the number of sharp peaks is larger for
GALs.  A similar qualitative resemblance can be seen for the real
part of the polarization function: at low energies the real part plunges
deeply (passing zero), whereafter it asymptotically
approaches zero from the negative side.  Again, the GAL exhibits
much more fine structure.

Also shown in Fig. \ref{ImPi12_3_lowmu}
are the results for the gapped graphene model. We observe that the
results depend on the azimuthal angle when the full GAL dispersion
is used, whereas the gapped graphene model does now show such
dependence. The overall shape of the gapped graphene model mimics
fairly well the GAL results, but is obviously much smoother
since the gapped graphene model with its only two bands cannot
reflect the van Hove structure of the full GAL dispersion.

While the gapped graphene model performs reasonably well at low
chemical potentials, as expected, it fails even qualitatively at
higher $\mu$.  An example is given in Fig.(\ref{ImPi12_3_highmu}),
where we show the imaginary part of the polarization function for
$\mu/t=0.5$, and, as can be seen, there is only a token of
resemblance between these results. The situation is even worse for
the real part (not shown).  We conclude that if one is
interested in the fine details of the polarization function, the
gapped graphene model is reliable only at very low chemical
potentials, and even there caution should be exercised.

\begin{figure}[t]
\includegraphics[width=8.5cm]{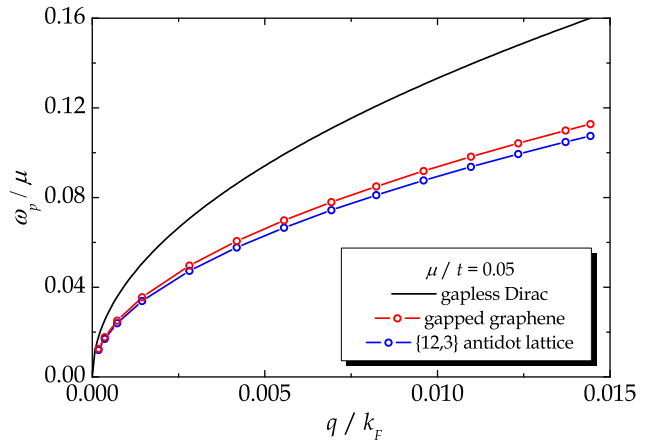}
\caption{(Color online) Plasmon dispersion for $\mu/t=0.05$.}\label{plasmon_lomu}
\end{figure}\begin{figure}[t]
\includegraphics[width=8.5cm]{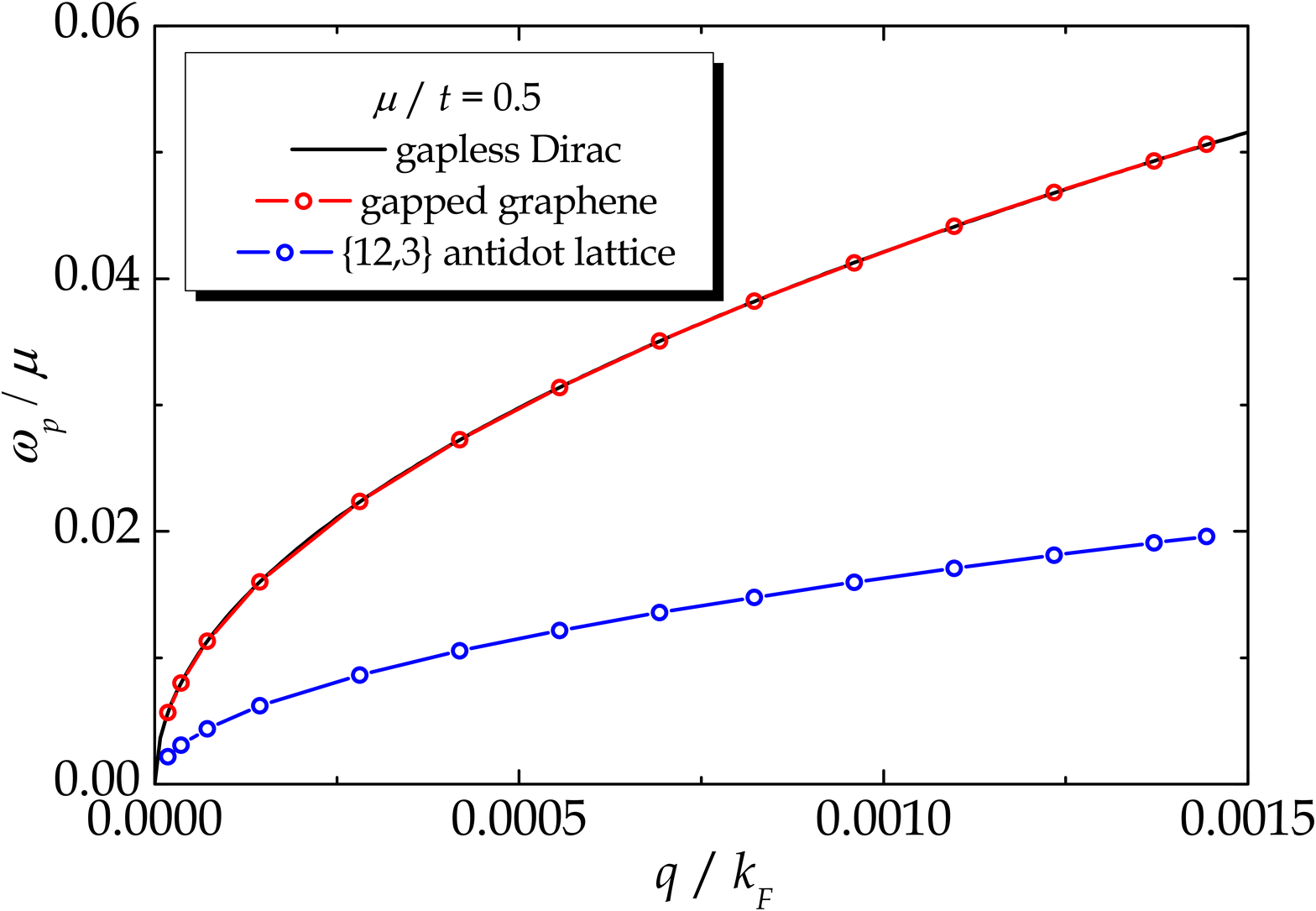}
\caption{(Color online) Plasmon dispersion for $\mu/t=0.5$.}\label{plasmon_himu}
\end{figure}

\section{\label{sec:plasmons}Plasmons}
We next examine the undamped plasmon modes for the $\{12,3\}$ GAL.
As usual, the modes $\omega({\bf q})$ are identified as solutions to
\begin{equation}
1-W({\bf q}){\rm Re}\chi_0^r({\bf q},\omega({\bf q}))=0,\label{plasmon}
\end{equation}
where $W({\bf q})$ is the Coulomb interaction.  We solve (\ref{plasmon}) numerically:
for a fixed ${\bf q}$ we find the $\omega_p({\bf q})$'s which cause Eq.(\ref{plasmon}) to vanish,
and use ${\rm Im}\chi({\bf q},\omega_p({\bf q}))=0$ as a criterion to identify the undamped modes.
In all cases considered the solution was unique.
We shall compare the numerically found GAL plasmons to the
well-known dispersions for pristine graphene\cite{Wunsch_NJP_2006}
$\omega_{{\rm gr}}$, and gapped graphene\cite{Pyat_JPCM_2009}
$\omega_{{\rm g-gr}}$:
\begin{equation}
\omega_{{\rm gr}}/\mu=C\sqrt{\frac{q}{k_F}},\quad \omega_{{\rm
g-gr}}/\mu=C\sqrt{1-\frac{\Delta^2}{\mu^2}}\sqrt{\frac{q}{k_F}},
\end{equation}
where $C=\sqrt{g_s g_v e^2/(8\pi\epsilon_r\epsilon_0 v_F)}$ with $g_s$
and $g_v$ spin- and valley degeneracies, respectively, and $\epsilon_r=2.5$,
the relative dielectric constant for graphene on a SiO$_2$ substrate. In the
general case the plasmon dispersion may depend on the direction of
the wave vector {\bf q}.  However, as can be seen in
Fig.~\ref{ImPi12_3_lowmu} (bottom panel), at large values of $\omega/t$ -- which
determine the undamped plasmons because ${\rm Im}\chi^r_0$ vanishes
there -- the azimuthal dependence is very weak, and can be
neglected. This result also follows analytically from the definition
of $\chi_0^r$, Eq.(\ref{chi0}): in the small $qa$ limit it is possible
to neglect the small energy difference $\epsilon_{n\bf k}-\epsilon_{n'\bf k'}$ as compared
to $\omega$, and the azimuth-independent $1/\omega$-behavior of Re$\chi_0^r$ follows, as
seen in the numerics of Fig. 3, bottom panel. In Figs. \ref{plasmon_lomu} and \ref{plasmon_himu} we
compare the plasmon dispersion laws for pristine graphene within the Dirac
cone approximation, a
gapped graphene with full $\phi({\bf k})$, and the $\{12,3\}$ GAL.
Qualitatively, the GAL plasmon dispersion always lies below the
other models.  At low chemical potential the gapped graphene plasmon
dispersion is essentially identical with the full calculation
(Fig.~\ref{plasmon_lomu}), while for a higher $\mu$ (Fig.~\ref{plasmon_himu}) the gapped
plasmon model works worse, as expected.
\section{Conclusions}
We have analyzed screening in graphene antidot lattices (GAL) within the random phase
approximation.  A general procedure for calculating the polarization function is
outlined.  An efficient long wavelength approximation is introduced, which eases
the numerical task considerably, and the accuracy of this approximation is analyzed
in terms of  numerical examples.  We also consider another phenomenological model: gapped graphene.
We have chosen a $\{12,3\}$ GAL as a generic system, and present results for this
structure evaluated within the various approximation schemes.  We conclude that the gapped
graphene model works reasonably well for low chemical potentials ($\mu/t\simeq 0.05$), but that it fails
even qualitatively for high chemical potentials ($\mu/t\simeq 0.5$).  We also determine
the undamped plasmons for the $\{12,3\}$ GAL.  The plasmon dispersion law for
the GAL has the same square-root
behavior as pristine graphene, but it is significantly suppressed.  The gapped graphene
model works very well for low dopings, because the high-energy tails of the real part
of the polarization are largely model independent.  Future issues that need to be addressed include
the effects due to the GAL symmetry, and the geometry and edge structure
of the nanoperforations defining the antidots.
\begin{acknowledgements}
APJ is grateful to the FiDiPro program of the Academy of Finland.
\end{acknowledgements}
{\it Note added in proof.}  During the technical processing of this manuscript we became aware
of very recent work by Scholz and Schliemann\cite{Scholz_PRB_2011}, who give a very detailed account
of the current-current correlation function of gapped graphene.  Our results agree with theirs, whenever
there is overlap.

\end{document}